\documentclass[iop,twocolumn]{emulateapj}
\newcommand{\Msol}{M$_{\odot}$}
\newcommand{\nh}{N$_2$H$^+$\ }
\newcommand{\nhns}{N$_2$H$^+$}
\newcommand{\hcop}{HCO$^{+}$\ }
\newcommand{\hthircop}{H$^{13}$CO$^{+}$\ }
\newcommand{\hthircn}{H$^{13}$CN\ }
\newcommand{\hcopns}{HCO$^{+}$}
\newcommand{\hthircopns}{H$^{13}$CO$^{+}$}
\newcommand{\hthircnns}{H$^{13}$CN}
\newcommand{\nhtwod}{NH$_2$D\ }
\newcommand{\nhtwodns}{NH$_2$D}
\newcommand{\mum}{$\mu m$}

\begin{document}
\title{Filamentary Accretion Flows in the Embedded Serpens South Protocluster}
\author{Helen Kirk\altaffilmark{1,2}, 
Philip C. Myers\altaffilmark{1},
Tyler L. Bourke\altaffilmark{1},
Robert A. Gutermuth\altaffilmark{3},
Abigail Hedden\altaffilmark{4},
Grant W. Wilson\altaffilmark{3}}
\altaffiltext{1}{Radio and Geoastronomy Division, Harvard Smithsonian
       Center for Astrophysics, MS-42, Cambridge, MA, 02138, USA}
\altaffiltext{2}{Currently a Banting Fellow at the Origins Institute,
	McMaster University; kirkh@mcmaster.ca}
\altaffiltext{3}{Department of Astronomy, University of Massachusetts 
	Amherst}
\altaffiltext{4}{Army Research Labs, Maryland}

\begin{abstract}
One puzzle in understanding how stars form in clusters is the source
of mass -- is all of the mass in place before the first stars are born, or
is there an extended period when the cluster accretes material 
which can continuously
fuel the star formation process?  We use a multi-line spectral survey
of the southern filament associated with the Serpens South embedded
cluster-forming region in order to determine if mass is accreting from
the filament onto the cluster, and whether the accretion rate is
significant.  Our analysis suggests that 
material is flowing along the filament's long axis at a rate of
$\sim30$~\Msol~Myr$^{-1}$ (inferred from the \nh velocity gradient along
the filament), and radially contracting onto the filament
at $\sim130$~\Msol~Myr$^{-1}$ (inferred from HNC self-absorption).
These accretion rates are
sufficient to supply mass to the central cluster at a similar rate to 
the current star formation rate in the cluster.  Filamentary accretion
flows may therefore be very important in the ongoing evolution of this
cluster.
\end{abstract}

\section{INTRODUCTION}
Understanding how clusters of stars form is a topic of considerable 
interest, since most stars appear to form within clustered
environments \citep[e.g.,][]{Lada03,Bressert10}.  As noted in \citet{Myers11a},
young stellar clusters are tightly packed with stars in their 
centres; the initial gas mass density inferred to create these
stars is significantly higher than observed formation conditions.
One simple solution to this problem is to feed in additional mass
as the cluster is forming.  Analysis of
competitive accretion simulations indeed show that this can
occur \citep{Smith09}.  Observationally, young embedded cluster-forming
systems are often found to be associated with multiple filaments,
in a hub-spoke like geometry \citep[e.g.,][]{Myers09a}.
Recent {\it Herschel} observations have shown that filaments
are ubiquitous in star-forming regions, and furthermore that
prestellar cores and protostars tends to be associated only with those 
filaments which are gravitationally supercritical \citep{Andre10}.
Filaments therefore appear to be the prime
candidate for providing a continuous source of material to be 
accreted onto forming clusters.  Few observations yet exist, however,
to test this idea, as most have focussed on the column density
structure of filaments and clusters \citep[e.g.,][]{Arzoumanian11}.  

Several decades ago, low resolution CO observations showed that
some filaments have velocity gradients running along their length
\citep[on $\sim10$~pc scales, e.g.,][]{McCutcheon82,Dobashi92}.  Additionally,
some filaments had hints of a velocity gradient
across their short axis \citep[e.g.,][]{Uchida90,Dobashi92}.
Higher resolution observations of dense gas species, however, 
have been lacking until recently.
\citet{Hacar11} observed the low-mass star-forming region 
L1517 in Taurus, and showed it contains several coherent, 
roughly thermal filaments which are well-fit
by an isothermal cylinder model.  These filaments
have small modulations in their (column) density and velocity
profiles along the long axis, suggesting core accretion from
the surrounding filament material.  Hacar et al (in prep)
conducted a similar study of the filaments in B213 in Taurus,
finding evidence that the singular filament visible in column
density maps is actually composed of many distinct velocity 
components of gas, each of which has a coherent and thermal
velocity profile. 
In more massive filaments, spectral observations tend to
indicate much more dynamic gas motions.  \citet{Miettinen12}
made pointed observations of C$^{18}$O and C$^{17}$O dense 
clumps embedded in the filamentary IRDC G304.74+01.32 and
find strong signs of infall and strongly non-thermal 
linewidths.  If the non-thermal linewidths can be interpreted
as providing additional support, then the clumps and
filament may be in virial equilibrium. 
In the DR21 filament, \citet{Schneider10} find signatures of
infalling motion over both the filament and clump scales,
and kinematic signatures consistent with the filament
forming from converging flows.

In order to study the question of filament accretion onto clusters,
we present kinematic observations of the Serpens South embedded cluster
and surrounding filaments.  To the best of our knowledge, this
is the first study to obtain sufficient complementary measurements
of the gas velocity to provide reasonable certainty in attributing
the motions to filamentary accretion.  

The Serpens South cluster was first discovered
in 2008 by \citeauthor{Gutermuth08} from {\it Spitzer} data.  In
this discovery paper, the cluster is clearly embedded in a dense
filamentary structure which shows up in absorption at 8~\mum.  
\citet{Gutermuth08} identify approximately 90 YSOs, and 
find that the ratio of class~II to I sources
in the region is low, around 0.7, which is much lower than the
median value of 3.7 in nearby cluster-forming regions
\citep{Gutermuth09}, and suggests that the Serpens South embedded cluster is
among the youngest examples known.  As such, Serpens South may
be a particularly good target to search for accretion flows in
filaments and onto the central cluster, since there will have
been less time for stellar feedback to disrupt or complicate
the cluster-forming gas motions.  Based on their {\it Spitzer} data,
\citet{Gutermuth08} estimate that stars are forming in the
central cluster at a rate of 90~\Msol~Myr$^{-1}$, although
more recent work by \citet{Maury11} using a combination of {\it Spitzer},
{\it Herschel}, and MAMBO 1.2~mm dust continuum 
observations suggest that the rate may
be several times smaller.  Much of this discrepancy is due
to the timescales assumed for protostellar evolution.  

Additional recent observations of Serpens South include near-IR
polarization measurements suggesting a relatively uniform
magnetic field aligned perpendicular to the main filamentary
structure \citep{Sugitani11}, and a catalog of protostars associated
with outflows suggesting that the energy injected in the central
embedded cluster by the outflows may be enough to maintain the
current level of non-thermal motion \citep{Nakamura11}.
X-ray observations from {\it Chandra} combined with the 
{\it Spitzer} data show that the YSOs have similar properties
to those in Serpens Main and NGC~1333 \citep[e.g,][]{Winston10}, 
with analysis suggesting that the class~III sources are
likely a by-product of early disk-stripping from inter-cluster
dynamics, rather than indicating a population older than
the class~II sources (Winston et al, in prep).

We obtained spectral line maps with Mopra\footnote{The Mopra
radio telescope is part of the Australia Telescope National Facility which 
is funded by the Commonwealth of Australia for operation as a National
Facility managed by CSIRO.} 
across $\sim$100~arcmin$^2$ of the main
filament in Serpens South in a variety of molecular tracers
(discussed in the next section) in order to measure the motion of the
dense gas in the filaments.  We combine this data with 
measurements of the column density distribution (Gutermuth et al, in prep)
obtained with ASTE using the AzTEC camera \citep{Wilson08} 
to fully interpret
the dynamical state of the filaments.  Using these data,
we find strong evidence that filamentary accretion flows
are playing an important role in the ongoing evolution of
the Serpens South cluster.

In Section~2, we discuss our Mopra observations, and the basic filamentary
structure seen in Section~3.  In Sections~4 and 5, we discuss analyze
the spectral data and estimate the filamentary accretion rates.  In
Section~6, we perform some simple analysis of the column density
structure of the filament, then discuss our main results in Section~7,
and conclude in Section~8.

\section{OBSERVATIONS}
\subsection{Data}
Our observations were made with the ATNF Mopra 22~m telescope 
in Australia, in September 2008, using
the single pixel broadband spectrometer, MOPS\footnote{The 
University of New South Wales Digital Filter Bank
used for the observations with the Mopra Telescope was provided with
support from the Australian Research Council}.  
MOPS allows for multiple spectral windows to be observed simultaneously,
increasing the mapping efficiency for multiple spectral lines;
the spectral resolution in the zoom mode is $\sim$0.108~km~s$^{-1}$
for the 90~GHz receiver.  
We used eight zoom windows to focus on different molecules of interest --
the (1--0) transitions of \nhns, HNC, HCN, \hthircnns, \hcopns, 
\hthircopns, plus SiO(2--1) and \nhtwodns(1,1). 
Most of these transitions trace dense gas; \hcop is 
sensitive to outflows, while SiO traces primarily shocked emission
\citep[see, e.g.,][for a discussion on the physical conditions traced
by these molecules]{Sanhueza12,BerginTafalla07,Vasyunina11}. 

We made on-the-fly maps, scanning along R.A., and Dec., with 11\arcsec\ spacing
between scan rows, and a scan speed of 3.71\arcsec~s$^{-1}$.  Square
maps of 5.3\arcmin\ per side (with the inner 5\arcmin\ fully sampled)
were made and later placed on a single grid using Gridzilla.
An OFF position of 
(18$^h$16$^m$56$^s$, +03$^o$02'19") 
was observed roughly every three minutes, and a measurement
of the system temperature taken every thirteen minutes.  Every map
required roughly an hour of integration time for each of the R.A. and
Dec. scans.
We used Livedata$^{3}$ to apply the bandpass calibration and
remove a low order polynomial baseline.
We then created uniformly gridded spectral cubes for each waveband
using Gridzilla\footnote{Livedata and Gridzilla are ATNF software 
for on-the-fly maps, written in AIPS++.  They are available at 
{\tt www.atnf.csiro.au/computing/software/livedata/index.html}.},
using a final pixel size of 15\arcsec; Mopra's spatial
resolution is $\sim$40\arcsec\ at 3~mm 
\citep{Ladd05}.
We converted the spectra from T$_a^*$ to T$_{mb}$ units using
an antenna efficiency of 0.49 \citep{Ladd05}.
Due to the relatively large separation between the adopted OFF position
and our mapping area, with this initial reduction, we found some
spectra still showed significant ripples in their baseline.
In these cases, CLASS\footnote{CLASS is available at 
{\tt www.iram.fr/IRAMFR/GILDAS}.} was used to 
fit and remove a higher order polynomial baseline.

\begin{figure}[h]
\plotone{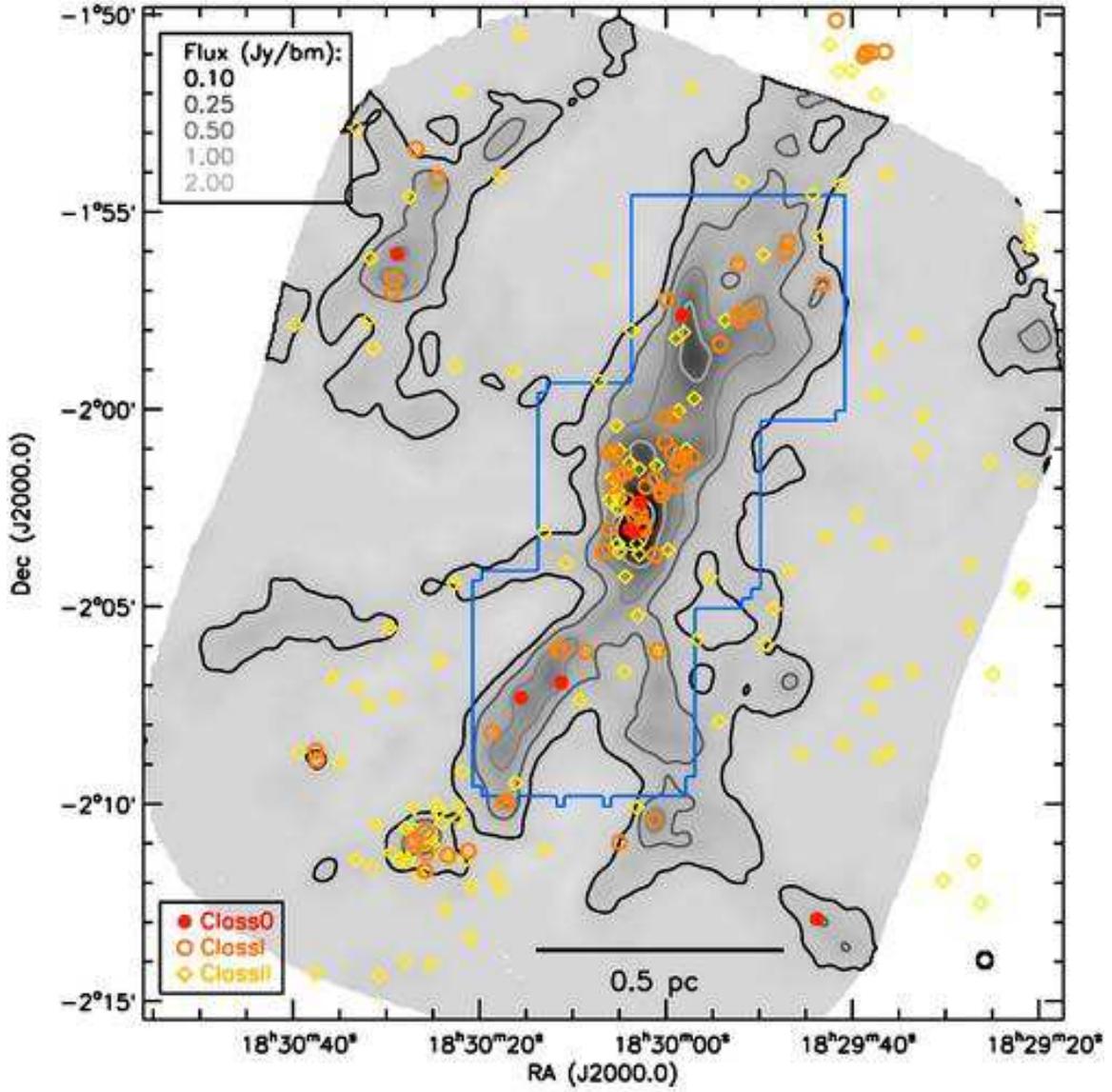}
\caption{The large-scale context of our Mopra observations of Serpens South.
	The greyscale image and contours show an AzTEC/ASTE map of the 
	millimetre 
	dust continuum emission, while the red filled circles, 
	orange open circles, and yellow diamonds
	show the YSO population found in {\it Spitzer} data. 
	The black circle at the bottom right corner shows the 
	AzTEC/ASTE beamsize (28\arcsec\ FWHM). 
	The blue contour shows the region
	mapped with Mopra, covering all of the main filamentary structure.}
\label{fig_context}
\end{figure}

Figure~\ref{fig_context} shows the region we mapped with Mopra,
compared to a dust emission map recently obtained with AzTEC/ASTE by
Gutermuth et al (in prep).  In Figure~\ref{fig_context}, the greyscale
image and contours show the dust emission, while the blue contour
shows the area we mapped with Mopra.  The yellow diamonds, orange
open circles, and red filled circles
show the YSOs identified by Gutermuth in a re-analysis
of the data in \citet{Gutermuth08} which now includes {\it Spitzer} 
24~$\mu$m data,
and adopts the classification scheme of \citet{Gutermuth09}.
Clearly, the Mopra observations
span the main filamentary structure in Serpens South.  In total, we
have more than 1500 spectra at each of the 8 wavebands, and cover
an area of roughly 100~arcmin$^2$.  The median value of the
rms noise per channel in the spectra are 
$\sim$0.34~K~km~s$^{-1}$ for all eight
molecular species.  Table~1 summarizes the 
central frequency adopted for each line and the associated reference.
For lines with hyperfine structure, the transition associated with the 
adopted frequency is also listed.

\subsection{Line Fitting}
For the emission lines without hyperfine structure (\hcopns, 
\hthircopns, and SiO)\footnote{\hthircop has been shown to have hyperfine 
structure by \citet{Schmid-Burgk04},
but the lines are spaced too closely to be discernable in our data, and so
we treat the line as having no hyperfine structure.},
we fit each spectrum
with a single Gaussian, using the least squares minimization code
MPFIT \citep[][see also {\tt http://purl.com/net/mpfit}]{Markwardt08} 
to constrain the fit parameters to
a reasonable range and calculate the errors associated with each fitting
parameter.  We similarly fit \nhtwod with a single Gaussian, as the line
was sufficiently weak that none of the hyperfine satellites were detected
(in most cases, not even the central component was visible).  
For these four transitions, the parameters fit are the velocity centroid, 
velocity dispersion (in Gaussian sigma units), and line peak.  

For the emission lines with hyperfine structure (\nhns, HNC, HCN, 
and \hthircnns), we substitute the Gaussian model with a single uniform
slab model with the appropriate hyperfine emission information
\citep[adapted from][]{Rosolowsky08}, and again use MPFIT to find the 
best fitting line parameters -- in this case, the velocity centroid,
velocity dispersion (in Gaussian sigma units), optical depth, and excitation
temperature.

For all of the results discussed in this paper, we only analyze
spectra which are above a signal to noise level of
3.5 -- i.e., we require an integrated intensity to be 3.5 times that
expected from the noise level (the spectral rms times the square root
of the number of velocity channels), and also a minimum of two spectral
channels showing emission about 3.5 times the spectral rms (using two
channels instead of one decreases the likelihood of the emission being
a noise spike).  Calculation of the integrated intensity is discussed
in Section~3.

\begin{figure}
\includegraphics[height=7in]{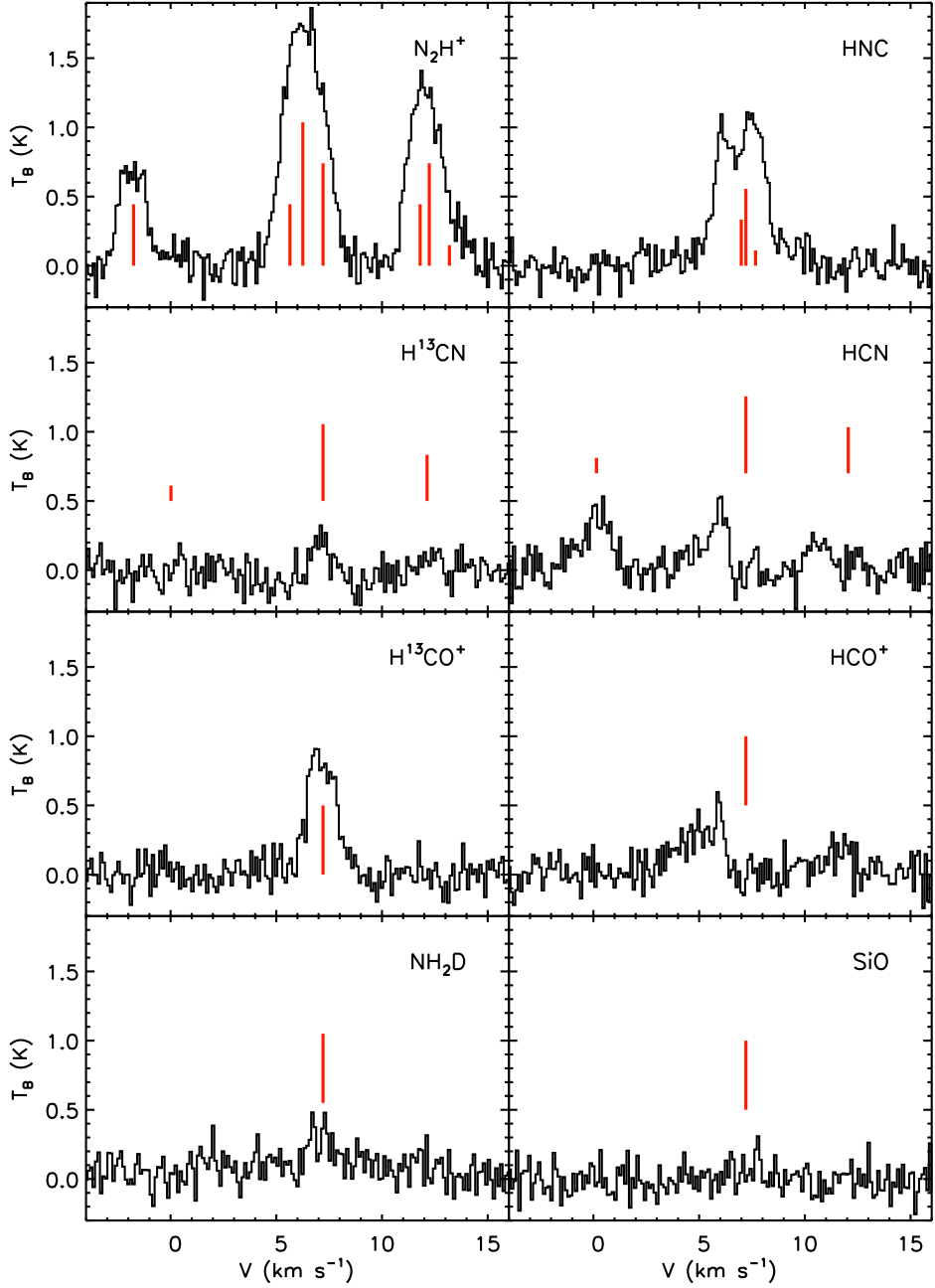}
\caption{The observed emission lines near the cluster centre  
	($-18^h30^m01^s, -2^o03'12"$).  The data have been summed
	over 4x4 pixels (an area of 60 by 60\arcsec) to increase the
	signal to noise level. 
	The red vertical lines indicate the expected line centres 
	based on a fit to the \nh spectrum.  In the case of 
	(discernable) hyperfine
	components, all are shown, with the relative heights scaling
	with the optically thin line ratios.  Note that
	SiO emission was not detected with certainty anywhere in the map.}  
\label{fig_central_bin}
\end{figure}

Over nearly the entire area mapped, most of the emission lines
detected show strong indications of self-absorption; only \nhns, \hthircopns,
and, where detected, \nhtwod appears single-peaked, while SiO 
has no unambiguous detections.  Even the \nh
lines may be mildly self-absorbed in some locations, but our
signal to noise and spectral resolution are insufficient to determine
this definitively.  
Figure~\ref{fig_central_bin} shows the spectrum of all
eight emission lines near the cluster centre, at the location where the
\nh emission is strongest, ($-18^h30^m01^s, -2^o03'12"$).  
In all cases, the spectra have been summed over the nearest
four by four pixels (60\arcsec) to increase the signal to
noise.  The red vertical lines indicate the location of the lines
based on the centroid velocity fit to the summed \nh spectrum.
Hyperfine components are indicated, along with their relative
intensities expected for the optically thin case.  As the figure clearly
illustrates, many of the emission lines have high levels
of self-absorption.
We therefore focus primarily on the line fit parameters for 
the \nh and \hthircop 
emission.  Note that since there 
are seven \nh hyperfine components with different optical depths, the 
\nh line widths and centroids fit tend to be satisfactory, 
even if a small amount of self-absorption is present.

\section{BASIC FILAMENT STRUCTURE: INTEGRATED INTENSITY}
\begin{figure}
\includegraphics[height=7.0in]{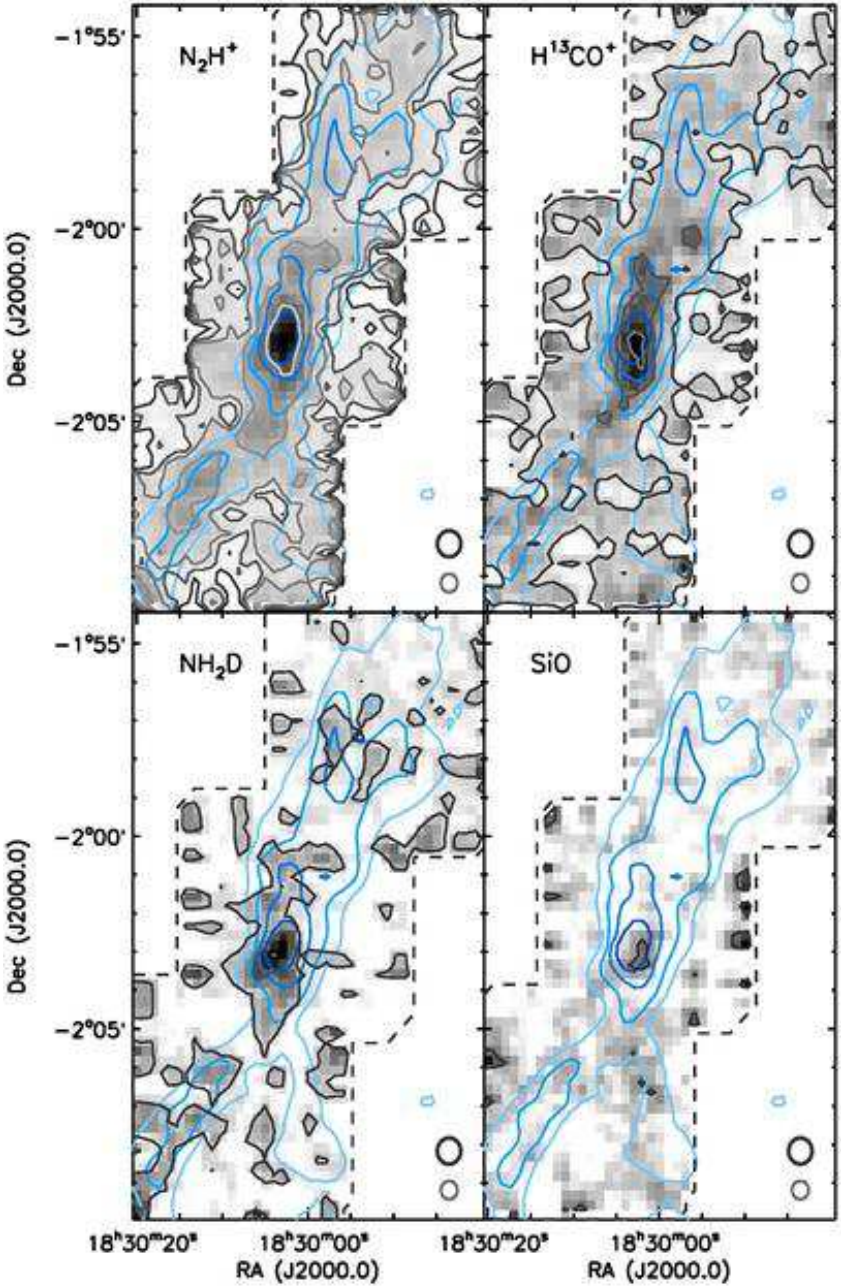}
\caption{The integrated intensity of the three brightest 
	molecular tracers in
	our survey not affected by self-absorption (\nhns, \hthircopns,
	\nhtwodns), as well as SiO, which should not show self-absorption,
	but has no clearly detected emission.
	The greyscale images show the integrated
	intensities, with white at a value of -0.1~K~km~s$^{-1}$ and
	black at a value of 90\% of the maximum
	integrated intensities measured in each molecular species
	(23.5, 3.8, 
	3.2, and 1.2~K~km~s$^{-1}$ respectively).
	The grey contours in all panels are at values of 
	0.5, 2, 3.5, 7, and 15~K~km~s$^{-1}$ from darkest to lightest.
	The dashed line indicates the full region mapped by Mopra,
	and the blue contours show the dust emission from AzTEC/ASTE,
	at levels of 0.25, 0.5, 1, and 2~Jy~bm$^{-1}$ from light
	to dark.  
	The Mopra and AzTEC/ASTE beams are shown on the bottom
	right of each panel in dark and light grey respectively.}
\label{fig_int}
\end{figure}

We first calculate the integrated intensity for the five 
molecular lines that do not exhibit self-absorption and hence are 
more likely to be optically thin
to show the column density structure of the dense
gas.  The full MOPS bandwidths are very large ($\sim$200~km~s$^{-1}$);
in order to minimize the contribution of noise to the calculation,
we restrict our integration to velocity channels close to where
emission is expected.  The velocity range is sufficient to extend
slightly beyond the maximum observed line extent; the range
adopted for each emission line is given in Table~1.
The velocity centre for integration
is, in order of preference, as the signal to noise level allows:
the centroid velocity fit to the particular line, the centroid velocity
fit to the \nh line\footnote{N.B.: the \nh line is the brightest and
most frequently detected of the five species without self-absorption.}, 
or the approximate
cloud-wide centroid velocity, 7.5~km~s$^{-1}$. 
Figure~\ref{fig_int} shows the integrated intensity maps for
\nhns, \hthircopns, and \nhtwodns, 
i.e., the brightest species without self-absorption, along
with SiO, which lacks any clear detections.
The typical uncertainty is given in Table~1.

Figure~\ref{fig_int} shows that both the \nh and \hthircop integrated
intensities have a similar structure to that seen in dust emission.
The integrated intensity for \nhtwod is much fainter, but appears to 
follow a similar trend.  All three of these molecular species are
expected to be good tracers of dense gas \citep[see, e.g., the 
discussion in][]{Sanhueza12,BerginTafalla07}, 
which is confirmed by
their good correspondence to the dust emission.  
\hthircn (not shown in Figure~\ref{fig_int})  
is also expected to be a good dense gas tracer 
\citep[e.g.,][]{HilyBlant10},
however, our data has too poor a signal to noise level to
see this definitively.  Summing our data over many pixels shows
hints that the central filament is brighter than the surrounding
gas, although even this is only seen at a marginal level.  
The \hthircn spectrum shown in Figure~\ref{fig_central_bin} is
one of the brightest spectra resulting from summing the data
over four by four pixels squared.

SiO is not detected with certainty anywhere in the map.  
As a tracer of shocked
gas \citep[see e.g.,][and references therein]{Caselli97}, 
its emission structure could be expected to differ from the
other four species.  The general lack of emission may indicate
a lack of strong (highly supersonic) shocks, however, with the 
noise levels in our observations, we cannot rule out weaker shocks.

\subsection{Filament Geometry}
In Figures~\ref{fig_context} and \ref{fig_int}, differing filament
shapes or geometries are visible north and south of the main cluster.
The southern filament is dominated by a single nearly straight filament
extending to the southeast, with several fainter filaments (e.g.,
pointing due south) apparent primarily in the dust emission map.
In the north, however, the filamentary structure is more complex.
A careful visual examination of both the
dust emission and molecular data to the northwest of the
main cluster shows a seemingly complex structure suggestive of 
overlap along the projected line of sight.
Intertwined filaments have recently been identified through
a careful analysis of the quiescent B213 filament in Taurus
(Hacar et al in prep). Other observations of filaments,
such as Orion \citep{JohnstoneBally99,Bally87} show morphologies
suggestive of similar phenomena in higher-mass systems.
Unlike B213, the large linewidths in the northern filament
of Serpens South make the separation of velocity components difficult
if not impossible, and for this paper, we restrict our analysis
to the simpler southern filament.

\begin{figure}
\includegraphics[height=7.0in]{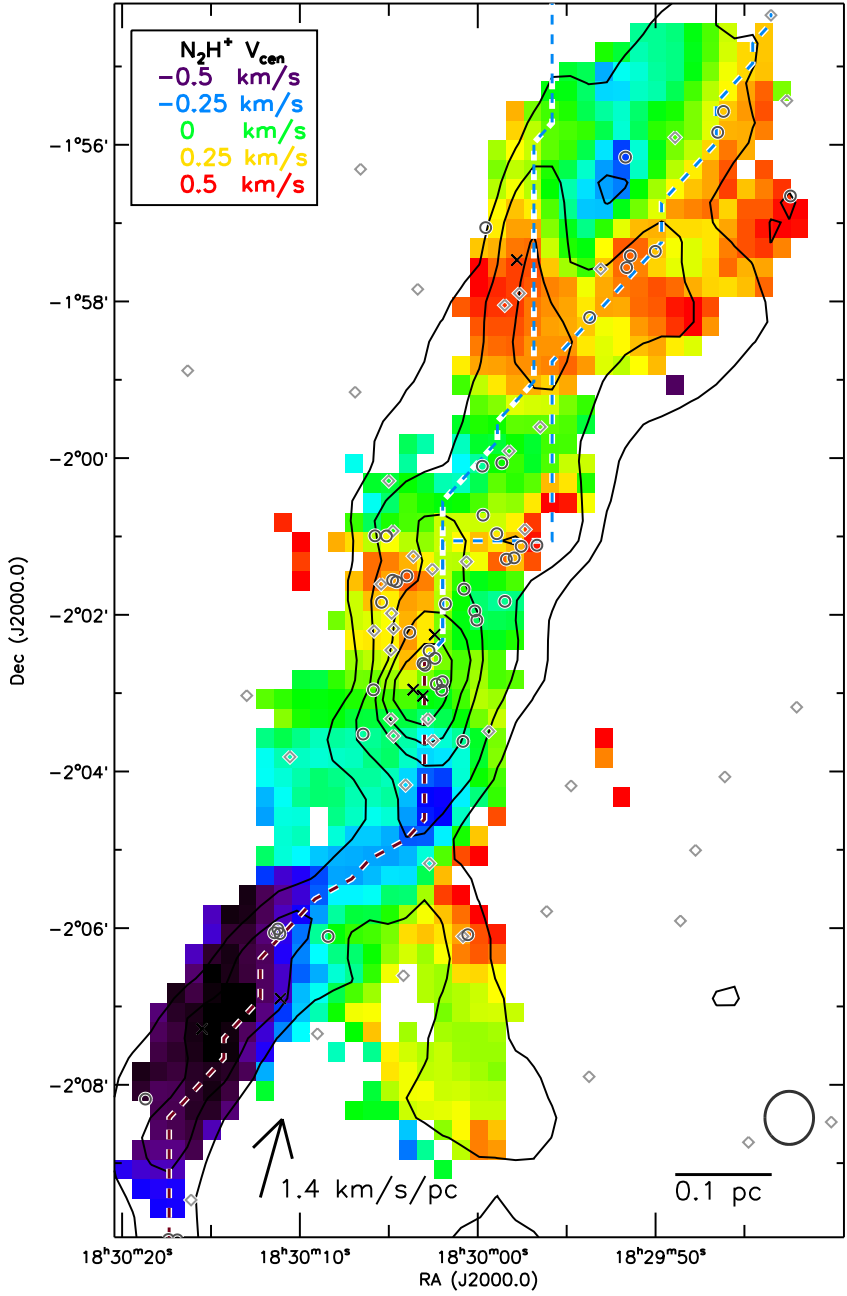}
\caption{The centroid velocity measured for \nh across Serpens South.
	The contours correspond to the dust emission levels shown
	in Figure~\ref{fig_context} and \ref{fig_int}, 
	and the circles denote the
	same YSOs as shown in Figure~\ref{fig_context} (dark grey
	corresponds to class 0, medium grey to class I and light grey
	to class II sources).  The peak ridge of the filament is
	shown by the dashed lines (red in the south and blue in the
	north for clarity).  The overall velocity gradient measured
	in the southern filament is indicated by the arrow at the 
	bottom left, while the circle at the bottom right indicates
	the Mopra beamsize.
	}
\label{fig_nh_glob_Vcen}
\end{figure}

We define the central or peak ridge of the southern filament by
searching for the location of peak emission along every horizontal
cut through the filament.  For this detection, we use the dust
continuum map; while the \nh integrated intensity map shows a 
similar distribution, the dust map has a higher signal to noise
level.  This filament definition is shown in 
Figure~\ref{fig_nh_glob_Vcen}.  For comparison, we also show a 
similar outline for the more complex northern filaments.
For our analysis of the southern filament, we consider the
data within 1\arcmin\ of the peak ridge (i.e., three 15\arcsec\
pixels on either side of the peak, or a diameter of
0.16~pc), and along a 0.33~pc extent.  
This area covers all of the
strongest emission present in the southern filament and excludes
any gas potentially associated with the much fainter filament
to the west.  Within this area, the southern filament 
has a mass of $\sim$20~\Msol, based on the dust emission
map\footnote{We assume a temperature of 15~K and a dust opacity of
0.0114~cm$^2$g$^{-1}$ \citep[e.g.][]{Enoch07}; recent ammonia
measurements suggest a dense gas temperature around 10-15~K in 
Serpens South, with the southern filament at $\sim$11-12~K
further from the central cluster and $\sim$12-14K approaching the
cluster (Friesen et al, in prep).}.
\citet{Maury11} estimate a mass
of 610~\Msol\ over a 15\arcmin\ by 25\arcmin\ area (i.e., several
times larger than the entire extent of our Mopra observations), 
based on 1.2~mm MAMBO
dust continuum data.
For both our estimate and that in \citet{Maury11}, a distance 
of 260~pc \citep[e.g.,][]{Straizys03}, is assumed, based on 
extinction estimates in the Aquila rift.  

There is, however, some debate over the true distance
to Serpens South.  Recent
VLBI measurements of a YSO in the Serpens main
core suggest that the distance to Serpens main
is roughly 415~pc \citep{Dzib10}.  Complicating
the picture is the fact that the gas velocities appear similar for
Serpens main, Serpens South, and Aquila; see the discussion
in \citet{Maury11} for more details.  Since most existing
analyses of Serpens South assume the closer distance of 260~pc
\citep[e.g.,][]{Gutermuth08,Maury11,Arzoumanian11},  
we also adopt that value here, but discuss in Section~7
how our results would differ if we instead assumed 415~pc.

\section{GLOBAL VELOCITY GRADIENT}
Figure~\ref{fig_nh_glob_Vcen} shows the centroid velocity fit to
our \nh spectra (where the signal to noise ratio is 3.5 or more). 
A clear velocity gradient is visible along the southern filament
(shown by the red dashed line), while the northern filaments
have a structured, but more complex velocity field.
If the southern filament is oriented with its southeast part
further from us and the cluster centre closer to us, then 
an accelerating accretion flow would produce a velocity
gradient similar to the one observed.  The infall line profiles
discussed in the next section lend further weight to this
choice of filament orientation.

We measure the velocity gradient for the southern filament following
the procedure outlined in \citet{Goodman93}.
We assume the \nh centroid velocities follow a simple linear form:
\begin{equation} A \Delta \alpha + B \Delta \delta = C
\end{equation}
where $\Delta \alpha$ and $\Delta \delta$ are the changes in right ascension
and declination across the map.  We found the best-fitting values to the
constants $A$ and $B$ using MPFIT \citep{Markwardt08}, 
and then converted all of the angular
velocity gradients to physical scale assuming a distance of 260~pc.
This yields an overall velocity gradient of 1.4 $\pm$ 0.2~km~s$^{-1}$pc$^{-1}$,
at an angle of $\sim 20^\circ$ west of north.  Note that while the 
\hthircop centroid velocities (not shown) 
agree quite well with the \nh centroid velocities,
fewer of the spectra have a signal to noise 
level above 3.5, which leads to a
less reliable measurement of the velocity gradient.

\begin{figure}
\includegraphics[height=5in]{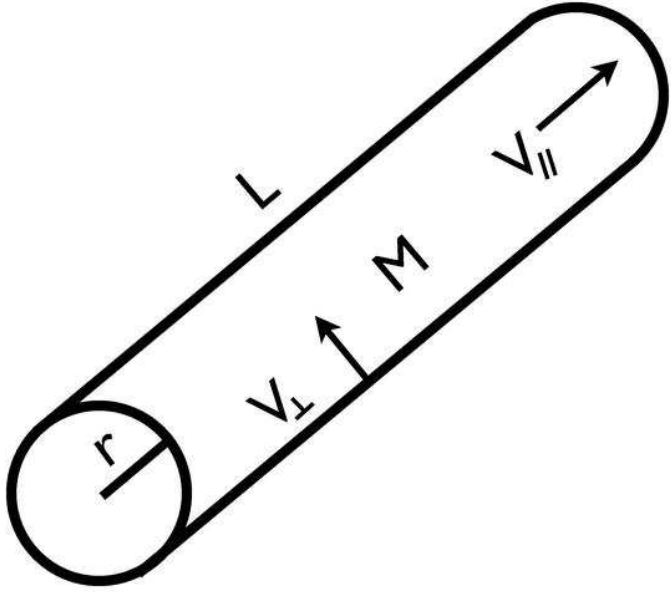}
\caption{The toy cylinder model we consider for our accretion rate
	analysis.  The cylinder has length $L$, radius $r$, mass $M$,
	and velocities of $V_{\parallel}$ along the long axis and
	$V_{\perp}$ in the radial direction.  It is inclined
	an angle of $\alpha$ (not shown) relative to the plane of
	the sky (with the top end further from the observer).}
	\label{fig_cyl}
\end{figure}

We can estimate the accretion rate implied by this velocity gradient
by using a simple cylindrical model, shown in Figure~\ref{fig_cyl}.
The cylinder has a mass $M$, length $L$, radius $r$,
an inclination to the plane of the sky of angle $\alpha$, and true
motions of velocities $V_{\parallel}$ along the filament long axis,
and $V_{\perp}$ infalling into the short axis (this final term will
be of interest in the following section).  The accretion rate,
$\dot{M}_{\parallel}$,
along the filament onto the central cluster is therefore
the velocity along the filament times the density and the area
perpendicular to the flow, or
\begin{equation}
\dot{M}_{\parallel} = V_{\parallel} \times \Big(\frac{M}{\pi r^2 L}\Big) (\pi r^2)
\end{equation}
which simplifies to
\begin{math}
\dot{M}_{\parallel} = V_{\parallel} \frac{M}{L}
\end{math}.
Due to projection effects, we observe
\begin{equation}
L_{obs} = L cos(\alpha) \quad  {\rm and} \quad 
V_{\parallel,obs} = V_{\parallel}\ sin(\alpha)
\end{equation}
with
\begin{math}
V_{\parallel,obs} = \nabla V_{\parallel,obs} L_{obs}
\end{math}.
Therefore, the accretion rate is given by
\begin{equation}
\dot{M}_{\parallel} = \frac{\nabla V_{\parallel,obs} M}{tan(\alpha)}
\end{equation}

For the southern filament, we estimate a mass of 20.3~\Msol\ along
a length of 0.33~pc and width of 0.08~pc (see Section~3.1), which 
corresponds to an accretion rate of
28~\Msol~Myr$^{-1}$ for $tan(\alpha) = 1$.  Note that since we observe
both a gradient along the filament and infall onto the filament
(discussed in the next section), this already constrains $\alpha$
not to lie close to either of the possible extreme values
(0$^{\circ}$ for lying parallel to the plane of the sky and
90$^{\circ}$ for lying directly along the line of sight).  
We will furthermore show that the accretion rates estimated from both the 
motion along and across the filament are relatively high, which
suggests that $\alpha$ is likely close enough to 45$^{\circ}$, so that
the correction factor for each measure is only several times unity.

The velocity gradient used
for the accretion rate estimate is also similar to values measured in
other star-forming regions -- for example, \citet{Bally87} find
a gradient of $\sim$0.7~km~s$^{-1}$~pc$^{-1}$ along several pc
of the Orion integral shaped filament; \citet{Schneider10} find 
gradients of $\sim$0.8-2.3~km~s$^{-1}$~pc$^{-1}$ along the DR21 filament,
and \citet{Kirk10} find a general relationship in non-filamentary
molecular gas structures between size and
velocity gradient suggesting values around 1~km~s$^{-1}$~pc$^{-1}$ for
size scales of a few tenths of a parsec.
The similarity of the measured gradient 
further suggests that projection effects are not extreme for
the filament, if Serpens South is not atypical.  

If the gradient were instead interpreted as rotation, then
the total rotational energy would be
\begin{equation}
E_{rot} = \frac{1}{2} \Big(\frac{M L^2}{3}\Big) \big(\nabla V\big)^2
\end{equation}
where the second term is the moment of inertia for a cylinder rotating
longways around its end, and the final term represents the angular velocity
squared.  For the southern filament, this yields
$E_{rot} \simeq 0.7$~\Msol~km~$^2$~s$^{-2}$.  The gravitational binding
energy is given by
\begin{equation}
E_{grav} = \frac{G M M_{clust}}{L}
\end{equation} where $M_{clust}$ is the mass of the central cluster,
roughly 100~\Msol, calculated in the same manner as the filament mass
(see Section~3.1).  Thus, the gravitational binding energy is a
factor of more than ten higher, at $E_{grav} \simeq 26$~\Msol~km$^2$~s$^{-2}$.
With a freefall time of just $\sim$0.3~Myr, 
several times less than the estimated age of the region, even if 
the filament was initially rotating, infall motions
would rapidly dominate.  

Finally, the velocity gradient {\it could} also be interpreted
as outflow motion, if the orientation was the reverse of that
assumed above.  In this instance, stellar winds and outflows would
be the most likely candidate source for driving the motions.
Using a simple calculation to estimate the
momentum and energy associated with such outflowing motions
yields values of 9~\Msol~km~s$^{-1}$ and 2~\Msol~km$^2$~s$^{-2}$ respectively.
\citet{Nakamura11} estimate that the total momentum and
energy in outflows (measured using CO(3-2) data) in Serpens South
is $\sim 8$~\Msol~km~s$^{-1}$ and $\sim 65$~\Msol~km$^2$~s$^{-1}$ 
respectively, with a total mass of outflowing material of only 0.6~\Msol.
While 
in principle, the YSO population could provide sufficient energy and
momentum to drive the observed large-scale flow, the outflows
observed by \citet{Nakamura11} are multi-directional, not
highly collimated, as would be required.

\section{INFALL ALONG THE LINE OF SIGHT}
We next utilize information from an optically thick line showing self-absorption
to measure gas accretion across (i.e., building up additional mass onto)
the filament.  In the classic case of two converging layers, the gas is 
assumed to have an increasing excitation temperature
toward the centre (where the density is also higher); self-absorption
in optically thick lines occurs
for the emission originating from the 
furthest pieces of gas at any given velocity.  Therefore, emission
from the (blue-shifted) layer moving toward the observer is emitted
primarily from the higher excitation, denser part of the slab, while the
reverse is true for the (red-shifted) layer moving away from the
observer \citep[e.g.][]{Myers96}.  Infall profiles therefore show brighter
peaks blue-ward of the peak emission of an optically thin line,
with a lower red shoulder, and sometimes a central dip.
Using simplified assumptions about the gas properties, the
emission profile of a line showing such a red-blue asymmetry
can be used to estimate a characteristic infall velocity of the gas
\citep[e.g.,][]{Evans99,Myers00}.

\begin{figure}
\plottwo{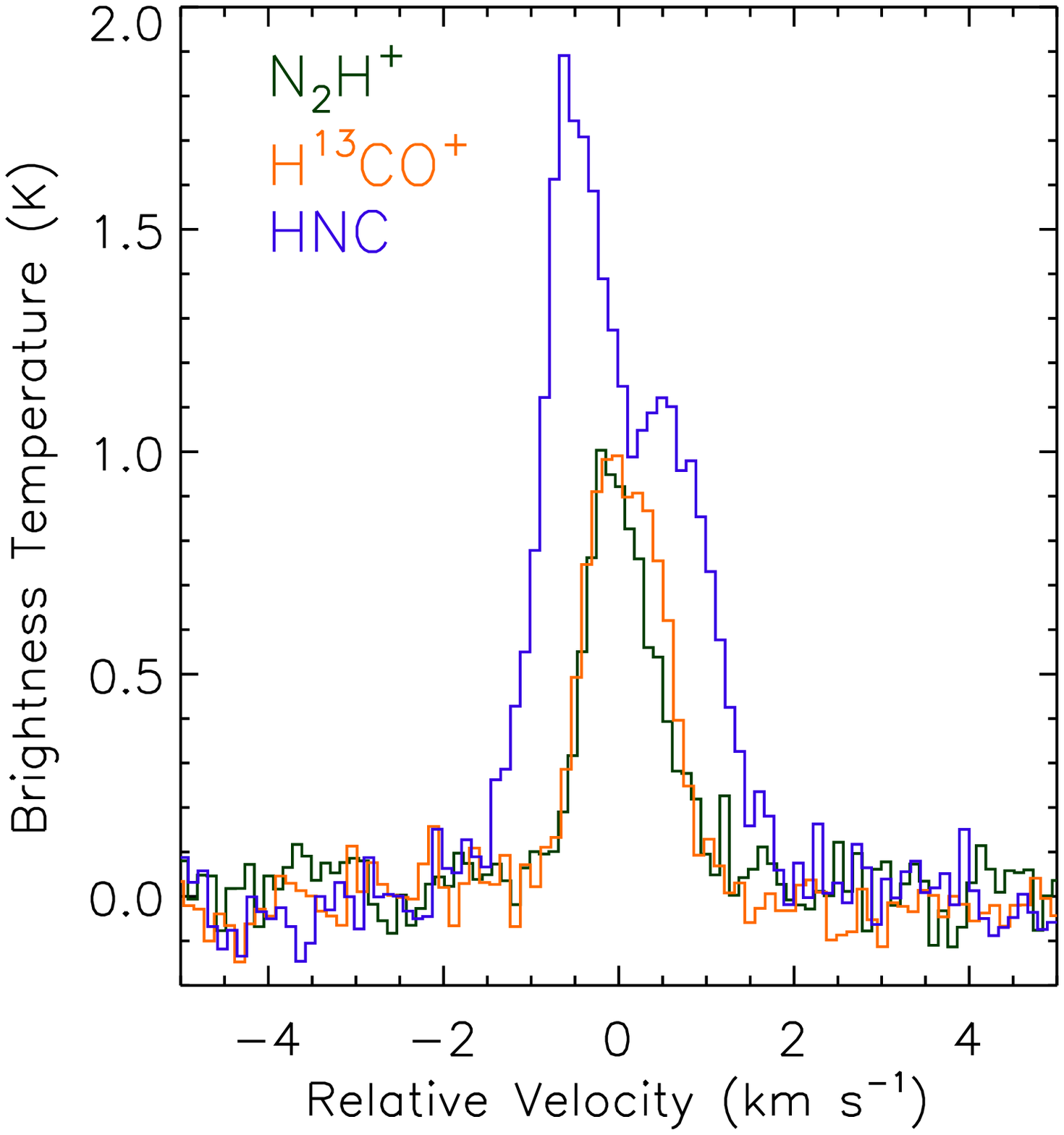}{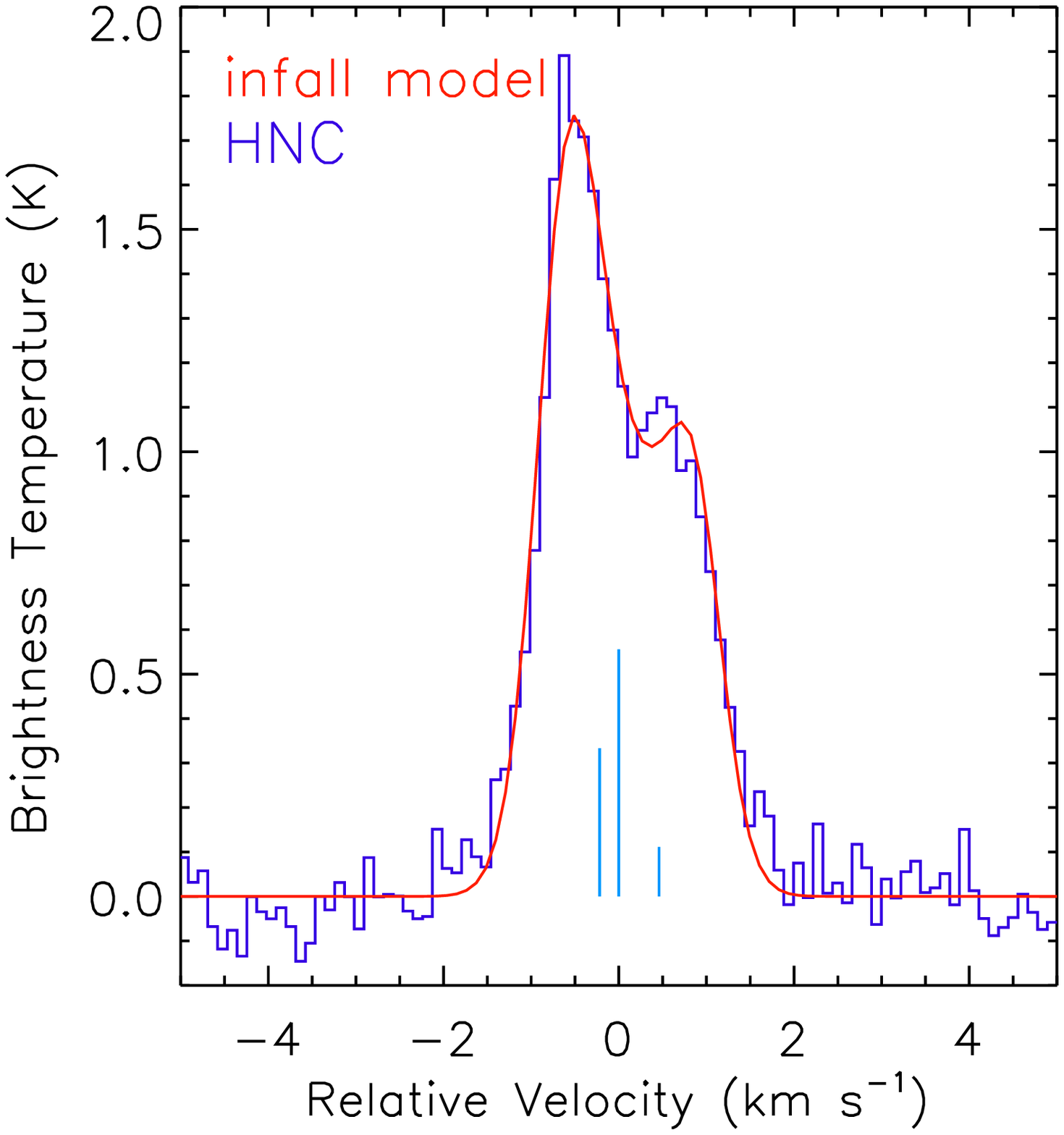}
\caption{Left: the \nh (green) and \hthircop (yellow), and HNC (blue)
	average spectra for the southern filament.  (For \nhns,
	only the isolated hyperfine component, (F$_1$,F=0,1-1,2), 
	is shown.)  
	The HNC spectrum shows clear signs of self-absorption, while
	the \nh and \hthircop spectra are well described by a single
	Gaussian.
	Right: The filament-averaged HNC profile, along with the
	expected relative positions and strengths of the hyperfine
	emission lines of HNC.  The hyperfine splitting is on too
	small a scale to be responsible for the observed
	self-absorption profile.  The red line shows the best fit
	Hill5 infall model profile \citep{deVries05}.  See text for details.}
\label{fig_HNC_vs_thin}
\end{figure}

In Serpens South, we observe self-absorption in several of our emission lines:
\hcopns, HCN, and HNC.  Of these lines, the HNC is the least self-absorbed,
and therefore is the brightest and has the highest signal to noise level.
A high signal to noise level is required to obtain reasonable model fits 
to infall spectra 
-- \citet{deVries05} find signal to noise levels $>30$ are required for good
results.
In order to further maximize our signal to noise level, we therefore spatially 
average spectra along the southern filament.
As discussed in the previous section, there is
a global velocity gradient running along the southern filament, which
needs to be accounted for prior to averaging.  We use the \nh centroid
velocities fit at each pixel to first shift all of the HNC spectra to a 
common velocity centre prior to averaging, again considering
spectra within 1\arcmin\ (or 3 pixels on either side)
of the filament's peak ridge identified.
Figure~\ref{fig_HNC_vs_thin} shows the result of this summation for
HNC, as well as for the two optically thin lines, \nh and \hthircopns.
In the case of \nhns, only the isolated component is shown for simplicity.
The HNC spectrum shows a brighter blue and fainter red peak with 
strong central dip around the peak emission of \nh and \hthircopns,
in agreement with expectations for infall.  The HNC(1--0) transition
does have hyperfine structure, however, this is on a much smaller
scale than the infall profile seen, as shown in the right hand
panel of Figure~\ref{fig_HNC_vs_thin}.

We estimate the infall velocity using the Hill5 model (and line fitting
code) described in \citet{deVries05}.  The Hill5 model considers gas
with a linearly increasing excitation temperature toward the centre,
with each half of the gas moving toward the other at a fixed
velocity.  
In their analysis, \citet{deVries05} found that the 
Hill5 model tended to provide the best fit to a range of both
simulated and real spectra with infall motions; the inclusion of
an increase in the excitation temperature allowed for much better
match than the classic two-slab model of \citet{Myers96}.  
The variables in the Hill5 model are the infall velocity, $V_{in}$,
the (Gaussian sigma) velocity dispersion, 
$\delta V$, total optical depth, $\tau$,
system velocity $V_{lsr}$, and peak excitation temperature, $T_{peak}$.
Note that the system velocity is expected to be 0 in our case, 
since the individual
HNC spectra were all shifted by the \nh centroid velocity fit at
the same pixel prior to the summation. 
Using
the Hill5 model, we find the best-fitting parameters listed 
in Table~2 under `Fit-1'.

In order to estimate the fitting errors, we ran the fitter on
1000 realizations of the spectrum with random noise added at the
same level as observed\footnote{Note that this increases the total
noise in the spectrum by a factor of $\sqrt{2}$.}.  
This test yielded several unexpected results.
In a small number of cases (26 of 1000), the `best' fit clearly 
got `stuck' in a local minimum of $\chi^2$ space, with unphysical
parameters (optical depths in excess of 20 and peak excitation
temperatures often
around several hundred).  These models provided visually poor fits to
the spectra, and so we discarded these cases.  These fits
likely could have been fixed by modifying the fitting tolerances in the
Hill5 fitter, but this seemed unnecessary, given the small overall
number.  Of the remaining 974
fits to the data plus random noise, 686 were found to have values 
clustered very close to the original fit (these were used to
calculate the formal errors in `Fit-1'), while the remaining 288
were clustered around a different set of fit parameters,
listed under `Fit-2' in Table~2.  Visually,
this other set of fits also are a reasonable match to the data, and
the distribution of $\chi^2$ values are similar for both sets
of fits.  Therefore, these second fit parameters appear to be 
a viable second solution to the best-fitting Hill5 model,
using only the information available from the HNC spectrum.

We use the additional information available from the optically thin 
spectra (\hthircop and possibly \nhns)
to argue that Fit-1 best matches the full suite of data.
We calculate the corresponding optically thin spectra for both
Fit-1 and Fit-2, using the fitted parameters but a lower optical
depth.  For Fit-2, the optically thin spectrum predicted has two distinct
velocity components separated by more than the linewidth.  Effectively,
Fit-2 represents a model of two independent layers of material moving
toward each other.  As shown in Figure~\ref{fig_HNC_vs_thin}, two 
distinct velocity components are not observed in either of the brightest
optically thin tracers, i.e., the \nh isolated
hyperfine component or \hthircopns.  This implies that Fit-1 is
the most consistent with the observations. 

We ran several simple tests with the online non-LTE radiative 
transfer calculator, Radex\footnote{Radex is available at 
{\tt http://home.strw.leidenuniv.nl/-}
{\tt$\sim$moldata/radex.html}.} 
\citep{VanderTak07}
to check that the fitted values are reasonable.  We assume
an HNC abundance of $6 \times 10^{-10}$ typical of dense core gas
\citep{HilyBlant10}, and adopt gas properties
(density, column density, temperature, total velocity dispersion) 
based on the
observations discussed earlier.  The predicted optical depth ranges 
from roughly 6 to 8, for temperatures (and the associated derived
densities and column densities) between 12 and 15~K.  Given that
a cold dense core HNC abundance is likely an overestimate for
a filament, and the uncertainties in various parameters used
to estimate the density and column density from the dust maps,
the Hill5 model fit of an optical depth of 2.7 for `Fit-1'
is in reasonable agreement; the optical depth of 1.6 from `Fit-2'
is less consistent with the observational expectations.

Using the same toy model as shown in Figure~\ref{fig_cyl}, the accretion rate
onto the filament can be estimated from the above infall velocity,
i.e., the velocity onto the filament times the density and filament
surface area of the short axes, or
\begin{equation}
\dot{M}_\perp = V_{\perp} \times \Big(\frac{M}{\pi r^2 L}\Big) (2\pi r L)
\end{equation}
which simplifies to $\dot{M}_\perp = 2 V_{\perp} \frac{M}{r}$.
Due to projection effects, we observe \begin{equation}
V_{in} = V_{\perp} cos(\alpha)
\end{equation}
while the observed filament radius / short axis should be invariant
under rotation.  Therefore, the accretion rate onto the filament
is given by \begin{equation}
\dot{M}_\perp = \frac{2 V_{in} M}{r cos(\alpha)}
\end{equation}
This corresponds to an accretion rate of 130~\Msol~Myr$^{-1}$ for
$cos(\alpha) = 1$ and is therefore a lower limit to the
true rate.  Additionally, the infall velocity fit with the
Hill5 model is expected to be a lower limit to the true characteristic
infall speed: the infall was modelled as two infalling parallel slabs,
while in a cylindrical geometry, some of the inward motions will
appear smaller along the line of sight.  Finally, note that 
using the alternative $V_{in}$ of
0.54~km~s$^{-1}$ from Fit-2 corresponds instead to a lower limit of 
$\sim280$~\Msol~Myr$^{-1}$.

Both the accretion rate onto the filament and from the filament
onto the cluster, without accounting for projection effects,
are within a factor of a few of the star formation rate
in the cluster estimated by \citet{Gutermuth09}, and somewhat
larger than the rate estimated by \citet{Maury11}.  Any 
correction for projection effects
made to the accretion onto the filament will serve
to further increase that estimate, i.e., beyond 130~\Msol~Myr$^{-1}$, 
while the accretion rate from the filament onto the cluster 
could be larger or smaller than our estimated 28~\Msol~Myr$^{-1}$
after correcting for projection.  

If the filament were projected at an angle of
30 to 60 degrees relative to the plane of the sky, then the 
accretion onto the central cluster measured would be within 
a factor of 0.6 to 1.7 of its true value, while the accretion onto the
filament measured would be a factor of 1.15 to 2 of its true value.  Based
on the combination of filament appearance, velocities measured,
and accretion rates derived, we believe that $30^{\circ} \leq \alpha \leq
60^{\circ}$ is a reasonable
range of filament angles.

\subsection{Other Lines}
\begin{figure}
\begin{tabular}{cc}
\includegraphics[width=8cm]{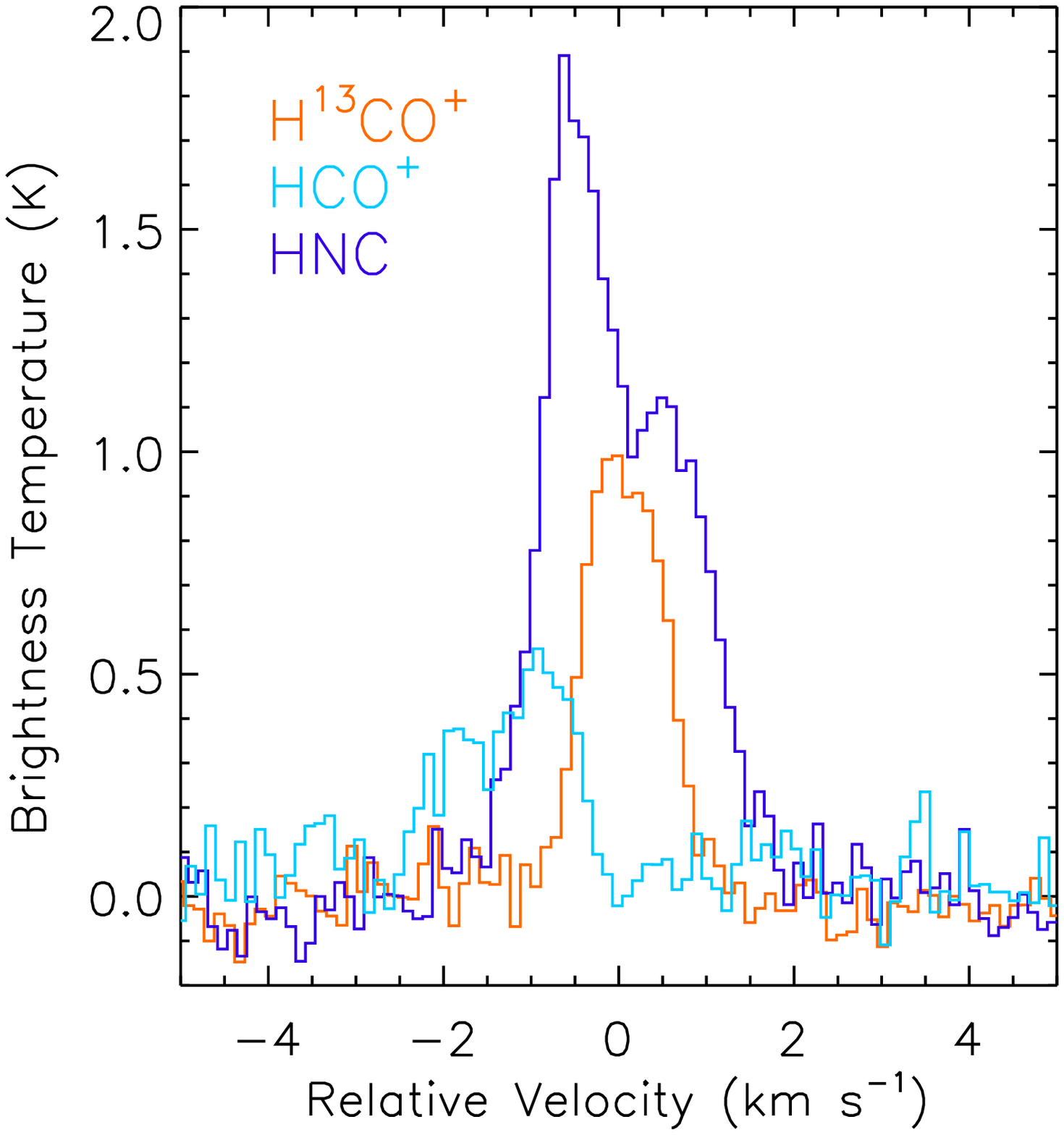} &
\includegraphics[width=8cm]{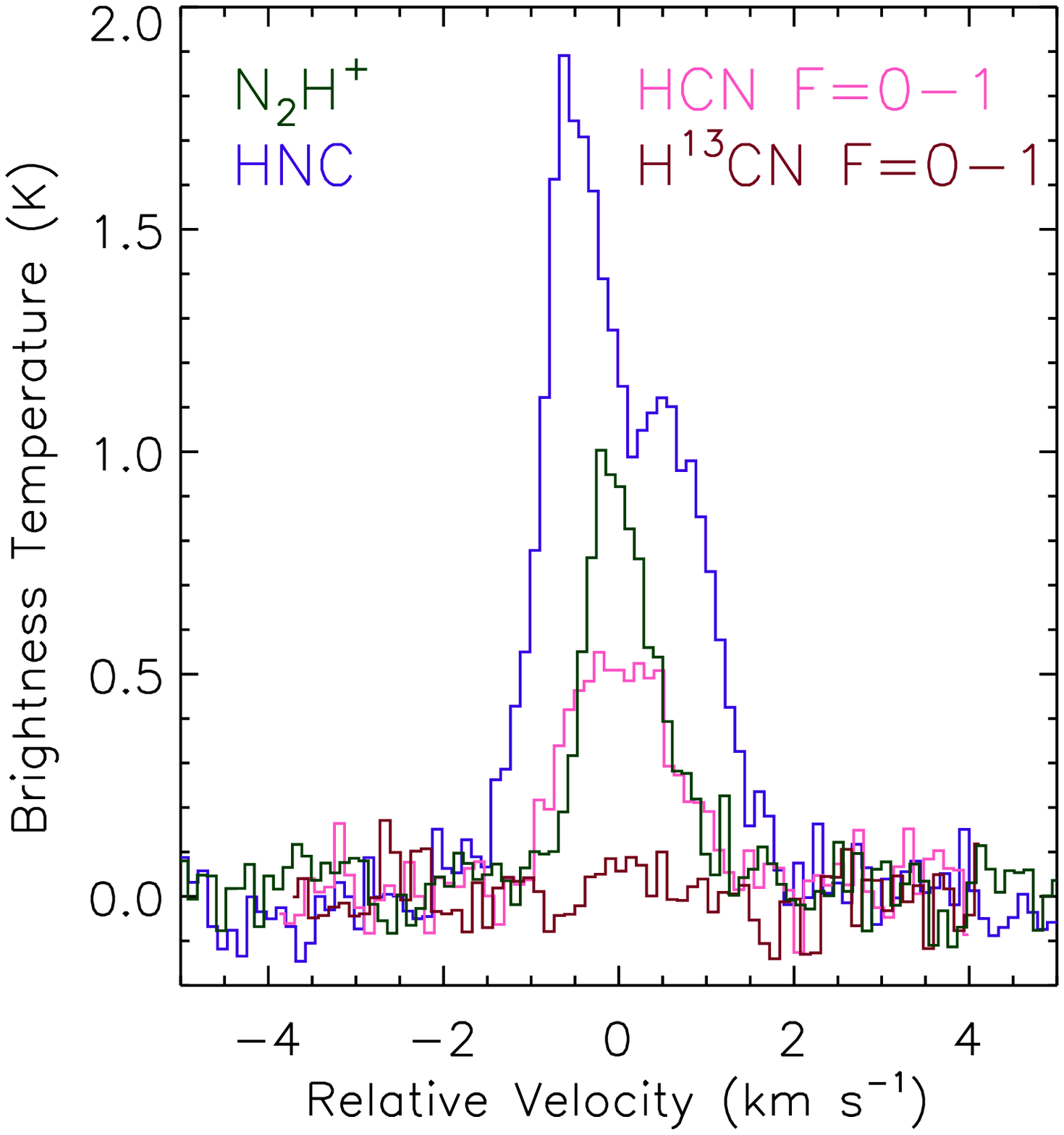} \\
\includegraphics[width=8cm]{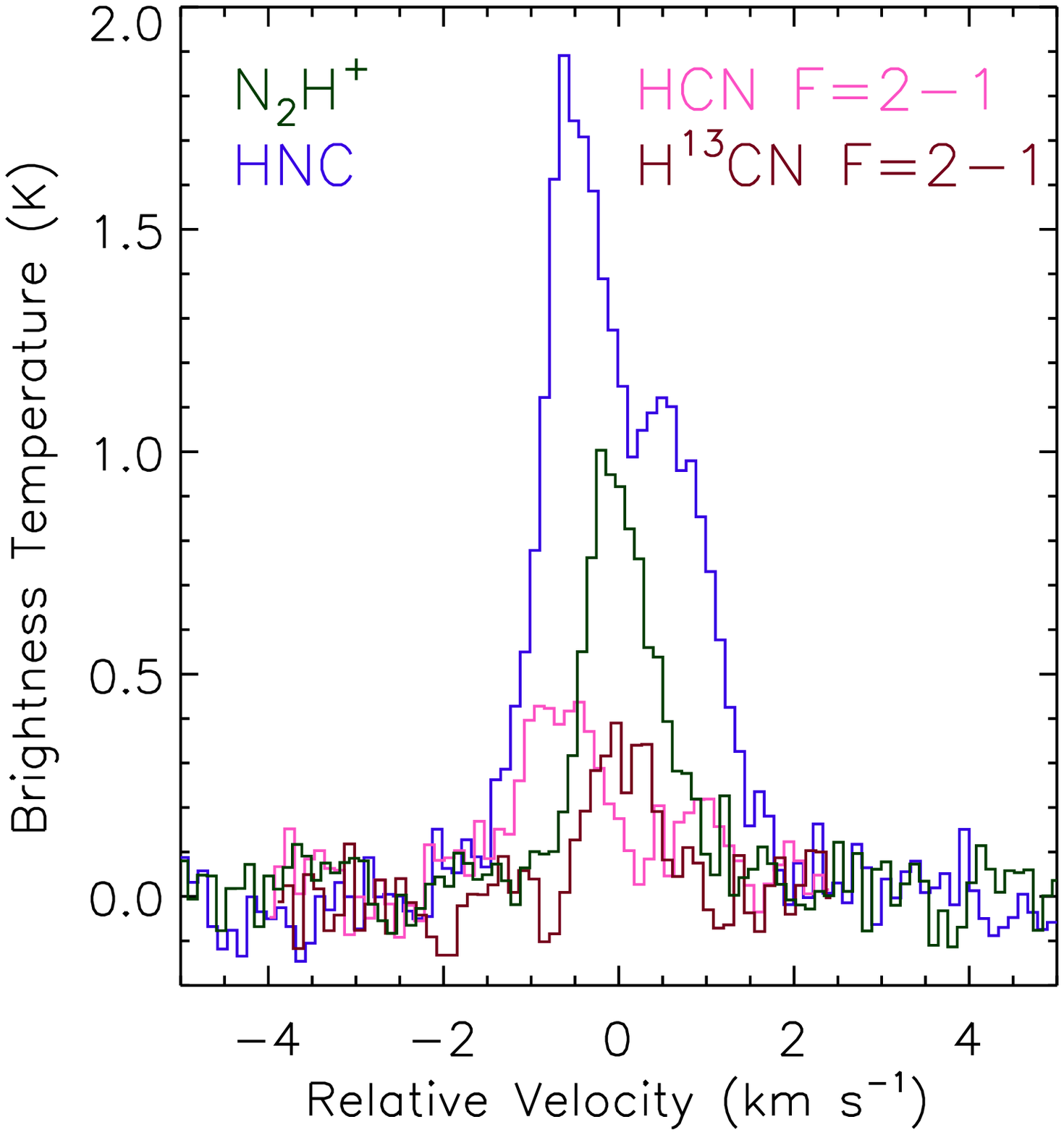} &
\includegraphics[width=8cm]{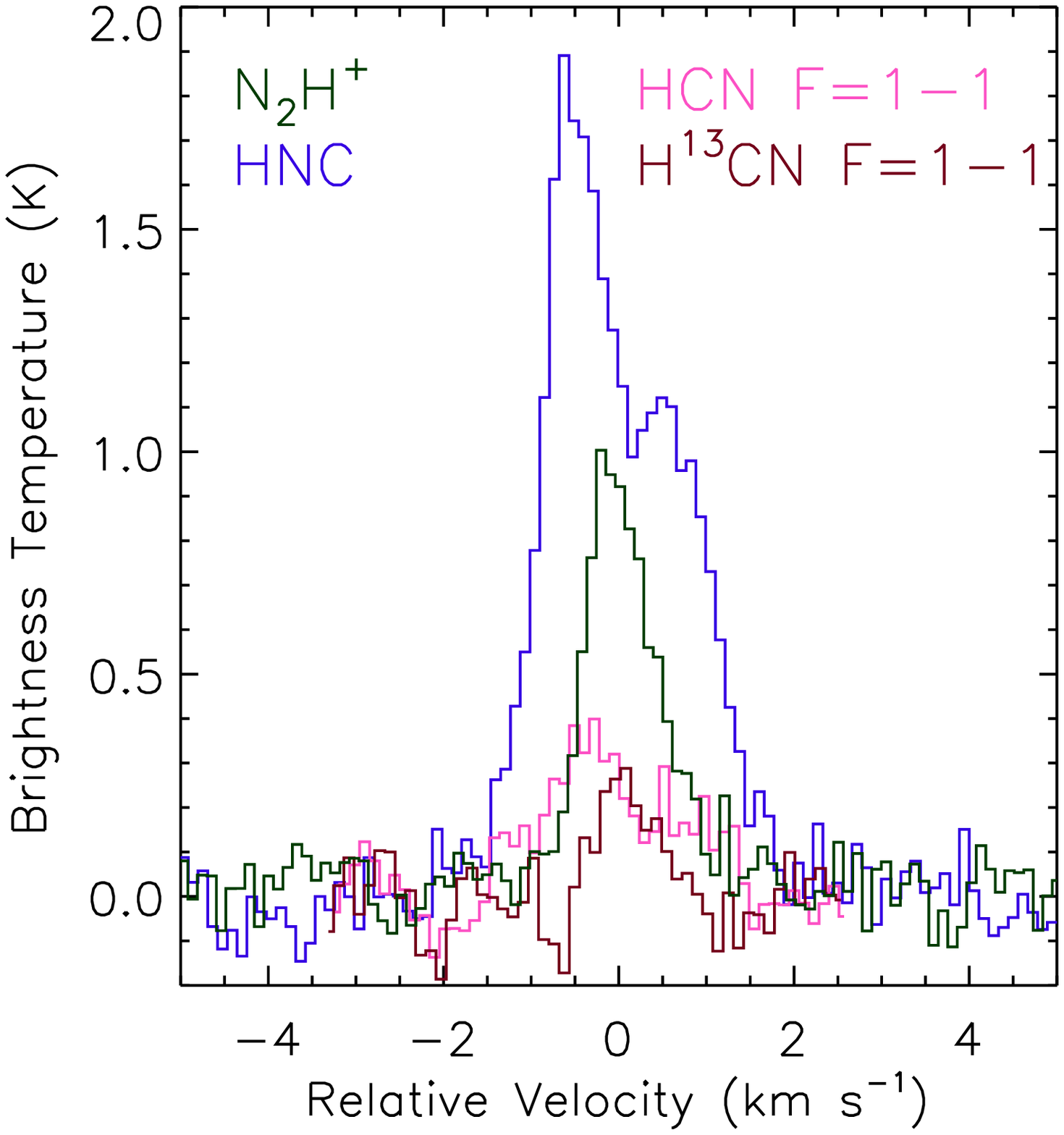} \\
\end{tabular}
\caption{A comparison of the self-absorption profiles for \hcopns, and the
	three hyperfine components of HCN.  The HNC emission is shown
	in dark blue, while \hthircop is in orange, the isolated
	hyperfine component of \nh (F$_1$,F=0,1-1,2)
	is in green, \hcop is shown in 
	light blue, and each component of HCN / \hthircn is shown
	in light / dark pink.  In the case of lines with hyperfine 
	components, the velocity centroid has been shifted to the
	appropriate common centre.  Note that no scaling has been
	applied to the brightness temperatures shown.}
\label{fig_infall_hcop}
\end{figure}

In addition to the self-absorbed HNC emission, both \hcop and
all three hyperfine components of HCN also show signs of self-absorption.
We performed the same spectral averaging across the southern filament for
these emission lines as described for HNC.  We additionally shift
each HCN (and associated \hthircnns) hyperfine component to the 
velocity it would have if it were the sole emission line.
Figure~\ref{fig_infall_hcop} shows the resulting spectral
line profiles.  As can be easily seen from the figure, all
four of the optically thick lines show such strong self-absorption
that relatively little emission is seen.  We attempted to 
model the emission lines with a Hill5 (or other) profile, but
due to the low signal to noise, even with averaging over the
filament, we were
unsuccessful.  

The degree of self-absorption in \hcop
and HCN is surprisingly large -- the \hcop emission peaks
at only half the value of the \hthircop peak, whereas
in an optically thin environment, the ratio would be somewhere
around a factor of 40 \citep[for non PDR molecular clouds at a 
galactocentric separation, similar to the Sun, of 8~kpc;][]{Savage02}. 
In \hthircnns, the line ratios
for optically thin gas would be 1:5:3 (for the top right,
bottom left, and bottom right panels in Figure~\ref{fig_infall_hcop}),
whereas HCN is clearly brightest in the (F=0--1) or first / leftmost 
hyperfine
component.  This would be consistent with a picture wherein the
highest levels of self-absorption occur in the optically thickest
components, and the least in the optically thinnest.

Recent work suggests that the rate coefficients for HCN and HNC
differ significantly \citep{Dumouchel10}, 
implying sensitivity to different density
ranges of material, and thus explaining why the two do not show
similar levels of self-absorption.
Alternatively, studies have shown that hyperfine
anomalies in HCN(1--0) are relatively common in higher column
density environments, and tend to lead to precisely the 
trend in peak brightnesses observed \citep{Loughnane12}.
The \hthircn emission is extremely faint, but appears
to be roughly consistent with optically thin emission with the predicted
line strengths (with no significant emission detected in the first / 
leftmost / F=0--1 component), this is also consistent with the
hyperfine anomaly study of \citet{Loughnane12}, who find that \hthircn
does not have anomalous ratios in sources where HCN does. 

We ran several additional checks to ensure that these severely
self-absorbed spectra are real, and were not affected by any
systematic effects.  For the velocity centroids,
all eight molecules were observed simultaneously with MOPS,
suggesting that any error with either the measured line velocities,
or the adopted central frequencies of each molecule, would
have to be present in all of our observations.  The Serpens South
data presented here are part of a larger project which includes
four other embedded cluster-forming regions.  Examining our
data in the other four regions shows very good agreement between
the different molecular line centroid velocities where there is
optically thin emission, suggesting that there are no systematic
effects with our velocity scale.  Similarly, the peak line
emission in these other regions, and the relative values of
peak intensity, agree well with expectations, suggesting no
systematic effects there either.  Finally, we note that similarly
severely self-absorbed emission line profiles showing a complete
absence of a red shoulder have also been observed in Serpens South 
with a different telescope and molecular line transition.
CO(3-2) emission observed with the ASTE 10~m telescope 
show a similarly shaped profile in some areas of the cluster
\citep[see, e.g., the bottom panels in Figure~6 of][]{Nakamura11}.

\section{COLUMN DENSITY PROFILE}

\subsection{Mass per Unit Length}
In this section, 
we examine some basic properties of the column density
structure of the southern filament.
The simplest property which can be measured for filaments is their
mass per unit length.  Assuming that a filament can be
approximated as an infinitely extended, self-gravitating, isothermal 
cylinder with no magnetic support, then the maximum
mass per unit length which can be in equilibrium is \begin{equation}
M_{line,crit} = 2 c_s^2 / G
\end{equation} 
\begin{equation}
M_{line,crit} = 16.7 \Big(\frac{T}{10~K}\Big)~M_{\odot}~pc^{-1}
\end{equation}
\citep[e.g.,][]{Ostriker64,Inutsuka97}.
Observations of the ammonia (1,1) and (2,2) inversion transitions
(Friesen et al, in prep) indicate that the kinetic temperature
across Serpens South is in the range of 10 to 15~K, implying
$M_{line,crit}$ is between 17 and 25~\Msol~pc$^{-1}$.
If the observed non-thermal motions were also considered to contribute
to supporting the filament \citep[e.g.,][]{Fiege00a,Miettinen12}, 
then with a typical velocity dispersion 
of $\sim 0.36$~km~s$^{-1}$ (Gaussian
sigma units) for \nh 
(mean or median value over the southern filament), 
then $M_{line,crit}$ is roughly 80~\Msol~pc$^{-1}$.

In the southern filament, we find a mass of 20.3~\Msol\ along 0.33~pc,
implying $M_{line} = 62$~\Msol~pc$^{-1}$, i.e., several times
the critical value, unless all of the non-thermal motion is
contributing thermal-like support.  
The total amount of support available is therefore likely
insufficient to keep the filament in equilibrium, implying that it
should be radially contracting.
Supercriticality is consistent with our velocity measurements
showing that gas is globally infalling in the filament. \citet{Arzoumanian11}
similarly find that filaments with higher column densities, such
as in Aquila, tend to have supercritical values of $M_{line}$.
Filaments associated with massive star formation tend to be
highly supercritical with $M_{line}$ often ten times larger 
than that in nearby regions \citep{Hennemann12}.

\subsection{Radial Column Density Profile}
In this section, we analyze the radial column density profile of the 
filament using the millimetre continuum emission.  We use
the peak ridge of the filament discussed in Section 3, and for each
pixel in the original AzTEC/ASTE data, we measure the column density
profile along lines of constant declination.  This provides for more
regular and better sampling than to measure points perpendicular to
the local filament orientation; we later correct our radial separation 
measurements to account for an average filament angle of 45$^{\circ}$.

All ground-based (sub)millimetre measurements are insensitive to the
largest scale structures, due to the observing methods which must
be employed.  For the AzTEC/ASTE map we analyze, Gutermuth et al (in prep)
have devised a novel technique which is able to recover flux out to
about 5\arcmin, a much larger scale than is typical.
In summary, the technique iteratively subtracts a model of 
the astronomical signal from the bolometer timestream data and then 
recomputes the atmospheric filtering and regenerates the signal model 
map.  The initial model is just the base pipeline-produced map masked 
to a signal to noise of 2, and convergence of the process is achieved 
when no new signal is found in an iteration down to a signal to noise of 3.
Despite this marked improvement in map reconstruction, 
the very largest-scale
structures with shallow or flat radial column
density profiles will not be included in the map.  
Some amount of large-scale structure is expected to be present, as
Serpens South is coincident with a roughly 
degree wide overdensity of dust extending to the
well-known W40 active star-forming region.
The 2MASS-based dust extinction map of \citet{Rowles09}\footnote{These 
maps are available at {\tt http://astro.kent.ac.uk/extinction}.}
show that
visual extinctions in excess of 5 to 10~mag are common across this area.
Some, but not all, of this extinction undoubtedly represents 
diffuse material unassociated with the individual star-forming regions
like Serpens South.  Any material that is associated with Serpens South,
however, would effectively result in a missing linear offset to the
derived column density profile; the Serpens South filament is sufficiently
compact that the {\it shape} of the column density profile should be
correct.
The function we fit
to the column density profile is dependent on absolute values, not
just the shape, so we limit our analysis to column densities where 
the fractional errors due to any missed large-scale structure are
reasonable.  
If a constant background of $A_V = 1$~mag is
missing from the AzTEC/ASTE map, there would be a
$\le$30\% underestimate of the inferred column density at 
fluxes of 0.067~Jy~bm$^{-1}$; we adopt this as our lower limit.
An additional complication to the analysis
arises from a second, fainter filament to
the west of the one we analyze.  In order to prevent this
structure from confusing the radial column density profile,
we additionally exclude pixels which lie westward of the minimum
value between the two filaments.

\begin{figure}
\plotone{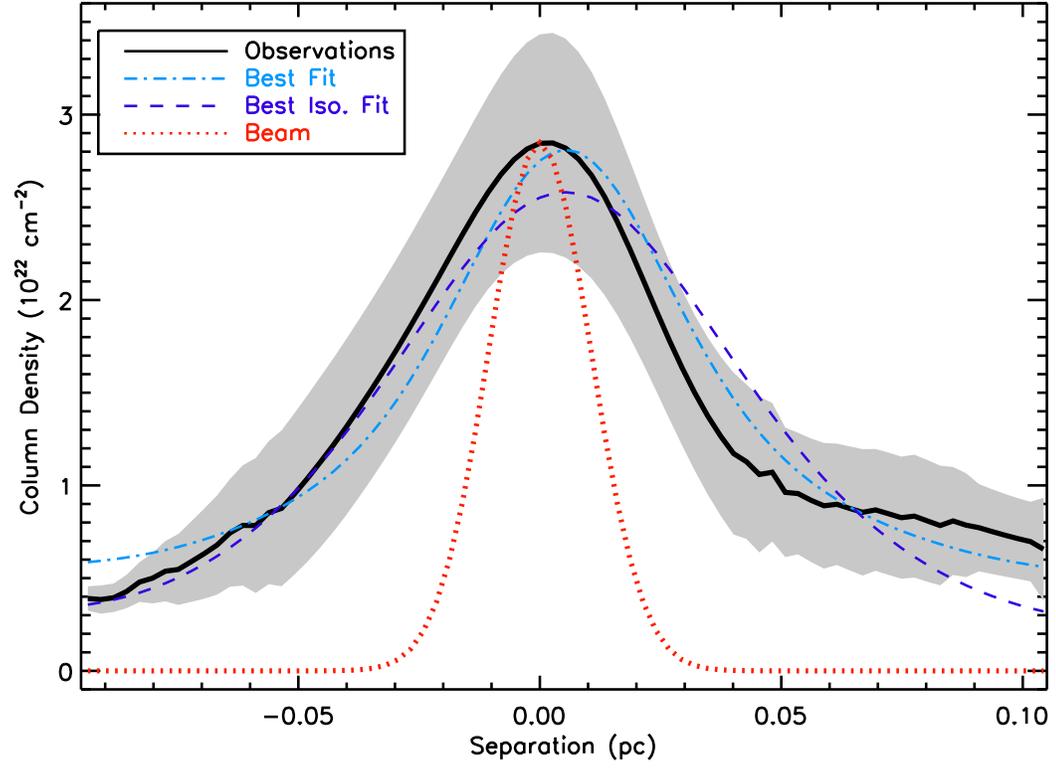}
\caption{The radial column density profile of the southern filament
	in Serpens South.  The solid black line shows the mean profile
	across the filament, while the grey shaded area 
	indicates the standard deviation.  
	The AzTEC/ASTE beamsize is shown by the red dotted line.
	The light blue dash-dotted line shows the best fit profile,
	following the formulation in \citet{Arzoumanian11}, while
	the dark blue dashed line indicates the best fit profile when
	the power law index is that corresponding to an isothermal 
	cylinder.}
\label{fig_radial_profile}
\end{figure}

After measuring the radial column density profile along the entire
peak ridge, we calculate the mean and standard deviation at 
each radial separation, similar to \citet{Arzoumanian11}.  
The mean profile and standard deviation of the column density
profile of the southern filament is shown in Figure~\ref{fig_radial_profile}. 
For comparison, the dotted red line shows the ASTE beam;
clearly the filament is resolved.

We compare the filament column density profile to the same suite
of models as in \citet{Arzoumanian11}: 
\begin{equation}
\Sigma_p(r) = A_p \frac{\rho_c R_{flat}}{[1 + (r/R_{flat})^2]^\frac{p-1}{2}}
\end{equation}
where $\Sigma_p$ is the column density at radial separation $r$, $\rho_c$
is the central (mass) density, $R_{flat}$ characterizes the inner flat
part of the profile, and 
\begin{equation}
A_p = \frac{1}{cos i} \int_{-\infty}^{\infty} \frac{du}{(1+u^2)^{p/2}}
\end{equation}
is a constant, and the inclination angle along the line of sight, $i$, is 
assumed to be 0 for simplicity.

Assuming that the observed column density profile is additionally convolved
with the ASTE beamsize, we use the MPFIT minimization code
\citep{Markwardt08} to determine
the best fitting parameters.  Allowing all three free parameters to vary,
we find $\rho_c = 3.6 \pm 2.2 \times10^5$~cm$^{-3}$, 
$R_{flat} = 0.010 \pm 0.006$~pc, and $p = 1.9 \pm 0.2$.  Our values for
$R_{flat}$ and $p$ lie within the range found by \citet{Arzoumanian11}
for all filaments within the Aquila Rift (note their $\rho_c$ values were
not published), including a power law index which is shallower than for
an isothermal cylinder.  It is important to note, however, that 
unlike \citet{Arzoumanian11}, we are not
able to make a strong a statement on the power law index of the
filament.  The restrictions discussed above
on where we are able to measure the radial column density profile 
limit our analysis to relatively close to the central ridge.
\citeauthor{Arzoumanian11}'s comparable radial column density
profile extends a factor of nearly 30 larger in radius than ours.
With the relatively small range in radial separations, our ability to
constrain the power law index is weak.  Figure~\ref{fig_radial_profile} 
shows also the
best fit obtained using a power law index of 4 (corresponding to an
isothermal cylinder), demonstrating that an isothermal model also
can provide a reasonable fit to our data.  The temperature 
required for the isothermal fit is 43~K,
which is similar to
the temperature implied by the non-thermal velocity dispersion in
\nh (around 50~K for $\sigma = 0.36$~km~s$^{-1}$).  This
is consistent with Section~6.1 where we showed that thermal pressure
was insufficient to balance the gravity.

Recently, \citet{Juvela12} created synthetic observations of filaments
formed in turbulent simulations, to investigate the effect of
instrumental noise and changing dust properties on the inferred
filament column density profile and mass per unit length,
especially as relates to {\it Herschel} data.  They find that
the mass per unit length is reasonably robust to these uncertainties,
while the parameters fit to the filament column density profile
can be more uncertain in cases of higher noise / poorer resolution
or a dust opacity which varies with density.

\section{DISCUSSION}
The southern filament in Serpens South has similar column density
properties to star-forming filaments recently studied with
{\it Herschel} \citep{Arzoumanian11}, with a mass per unit length
several times larger than the critical value for thermal
support, and a radial
column density profile which is consistent with $p \sim 1.5-2.5$.  
In fact, the Serpens South filament 
is included in the \citet{Arzoumanian11}
sample of `Aquila' filaments, although the individual filament
locations / properties are not given in their paper to allow a
more detailed comparison.  The relatively large {velocity dispersions} seen
in dense gas tracers in the southern filament could be interpreted
as evidence for providing sufficient stability against gravitational
collapse, however, our more detailed kinematic analysis suggests
that the filament is in fact not in equilibrium.  We find
evidence for mass flows both along the filament onto the
central cluster and onto the filament from the less dense
surroundings, with accretion rates of order 28 and 130~\Msol~Myr$^{-1}$
respectively.  These rates are uncertain by a factor of a few, for
several reasons, including the angle of the filament with respect
to the plane of the sky.

Although both accretion rates are uncertain
to a factor of a few, it appears likely that the 
accretion rate onto the filament is larger than the 
accretion rate from the filament onto the central cluster.
Formally, in order to have the latter rate be higher, the
filament angle would need to satisfy
\begin{equation}
\frac{28~M_{\odot}~Myr^{-1}}{tan(\alpha)} \ge 
\frac{130~M_{\odot}~Myr^{-1}}{cos(\alpha)}
\end{equation}
which is satisfied for $\alpha \le 12^{\circ}$, i.e.,
a filament which is nearly parallel to the plane of the sky.
Such a small inclination angle seems unlikely, but note that
uncertainties in all observed quantities
(mass, radius, velocity gradient, and infall velocity) could
allow for a slightly larger range of angles.  One must
always use caution in interpreting velocity measures from 
different molecular tracers.  The gas velocity
could be some (non-constant) function of density, and using
different molecular tracers for each accretion rate measure
could in principle reflect that, rather than overall flow
differences in the two directions. 
The only way to circumvent this potential issue entirely would
be to use observations of a single self-absorbed molecule and
its rarer isotopologue for both measures.  We do not have such
data available at high enough signal to noise, however, 
the consistency of the HNC self-absorption dip with the \nh and \hthircop
peak emission suggests all are tracing similar volumes of gas,
and that this should not be a major problem.

If the tentative difference in accretion rates is true, 
this could imply
that in the future, the accretion rate onto the central cluster
will also increase (given the larger amount of mass available in the
filament), and thus lead to more vigorous star formation in the
cluster centre.  It is also possible that the increase in 
mass in the filament could promote local instabilities in
the filament, leading to core formation, and eventually to a 
series of star formation sites along the filament.  
Dense cores are, in fact, already observed in the AzTEC/ASTE data
along the filament, as well as in 8~$\mu$m absorption.  A small
number of class 0 and I YSOs are also seen, which may be an
indication of past instabilities in the filament.  The small cluster
of class II YSOs just past the southern end of the southern
filament suggests a previous episode of star formation, perhaps
also associated with the filament.

At the current estimated accretion rate of 28~\Msol~Myr$^{-1}$,
the cluster's membership would roughly double in 1~Myr with the
addition of about 60 YSOs, assuming a mean mass
of 0.5~\Msol.
An increasing accretion rate would lead to a larger number of
YSOs.
From these estimates, we can see that our estimated accretion
rate from the filament onto the cluster is at a comparable 
level to previous observational estimates
of the rate of mass supply required to form the current generation
of YSOs in the cluster centre.  We can therefore conclude that
filamentary accretion is an important, if not the dominant,
factor in governing the formation and evolution of stars in the
central cluster.

Throughout this analysis, we have adopted a distance to Serpens
South of 260~pc \citep[e.g.,][]{Straizys03}, for consistency with
previous work, although recent VLBI observations
suggest that a distance of $\sim415$~pc \citep{Dzib10} applies
to at least the main Serpens cluster.  If this distance is also
appropriate for Serpens South, then the quantities we estimate
in our analysis here become a factor of several larger.
The mass in the southern filament would increase by a factor
of 2.5, the velocity gradient would decrease by a factor of 1.6,
and all other quantities of interest (lengths, accretion rates,
mass per unit length) increase by a factor of 1.6; the accretion
rates become 45~\Msol~Myr$^{-1}$ parallel to the filament (onto the
cluster) and 208~\Msol~Myr$^{-1}$ perpendicular to the filament.
None of these changes
effect our conclusions qualitatively, and in fact, they tend
to strengthen our finding that filamentary accretion is important
in Serpens South.

Filament-cluster systems such as Serpens South present a unique
opportunity to gain a deeper understanding of cluster formation.  With
more sensitive observations, it should be possible to measure the spatial
variation in the infall velocity across the filament, which could be
used to model the full 3D velocity profile of material in the filament.
This in turn can provide a straightforward test of the relative importance
of magnetic fields, gravity, and turbulence in the ongoing cluster
accretion, as discussed in \citet{Balsara01} and Heitsch (in prep). 

\section{CONCLUSION}
Using a combination of dust column density information derived from
AzTEC/ASTE continuum maps (Gutermuth et al, in prep) and dense gas
kinematics derived from Mopra spectral maps, we study the flow of
gas along the southern filament of the Serpens South embedded 
cluster.  We find that the filament appears typical in terms of
column density properties (e.g., a supercritical mass per unit
length).  The multiple molecular emission lines observed in our
Mopra survey allow us to infer the possible 
presence of an accretion flow from the filament onto the 
central cluster (if the filament is in front of the cluster), 
and additionally
we find that material continues to be accreted onto the filament.
We estimate accretion rates of 28~\Msol~Myr$^{-1}$
for accretion onto the central cluster, and a lower
limit of 130~\Msol~Myr$^{-1}$ for infall onto the filament
(assuming a distance to Serpens South of 260~pc). 
Both of these are similar in magnitude to the
estimated current rate of star formation in the central cluster.

Filamentary accretion flows therefore appear to be an important
mechanism for supplying the material necessary to form the
YSOs in this cluster.  As more embedded clusters and their
associated filaments are identified in current surveys, it will
be interesting to measure the associated gas kinematics to
determine whether other embedded clusters behave similarly to
Serpens South.

\acknowledgements{
We thank the referee for suggestions which helped to clarify aspects of
this paper.
We thank Steve Longmore for assistance with
the Mopra observing setup and data reduction.  We thank Phil Edwards at ATNF
for allocating additional time to complete our Mopra observations,
and all the ATNF Narrabri staff for their support.  
We thank Rachel Friesen for sending her ammonia temperature results
prior to publication, and for several interesting discussions
on Serpens South.  We thank Elaine
Winston and Scott Wolk for sharing the results of their {\it Chandra} survey
of Serpens South prior to publication.
HK thanks Paola Caselli for a valuable discussion
on astrochemistry, Chris de Vries for answering queries on his infall
fitter on numerous occasions, and Fabian Heitsch for a critical 
read-through of an earlier draft of this paper, as well as sending
a copy of his filamentary accretion paper pre-publication.  
HK acknowledges funding from a Smithsonian Scholarly 
Studies Program grant.
This research was supported in part by the National Science Foundation under
grant number 0708158 (TLB).  
}

\begin{deluxetable}{cccccc}
\tabletypesize{\normalsize}
\tablecolumns{6}
\tablecaption{Basic Line Properties \label{tab_molec_props} }
\tablehead{
\colhead{ } &
\colhead{Frequency} &
\colhead{HF component\tablenotemark{a}} &
\colhead{Ref.\tablenotemark{b}} & 
\colhead{Int. Range\tablenotemark{c}} &
\colhead{Integrated Noise\tablenotemark{d}}\\
\colhead{Molecule} &
\colhead{(MHz)} &
\colhead{(if applicable)} &
\colhead{ }  &
\colhead{(km~s$^{-1}$)} &
\colhead{(K~km~s$^{-1}$)} 
}
\startdata
N$_2$H$^+$(1--0) 	& 93173.4669 & $F_1,F=2-1,2-1$  & 1 & -10.5$\to$7.5 & 4.38\\
H$^{13}$CN(1--0) 	& 86340.1840  & $F=2-1$		& 2 & -8.5$\to$6.5  & 3.87\\
HCN(1--0)		& 88631.8473 & $F=2-1$		& 3 & -9$\to$7 & 4.02\\
HNC(1--0)		& 90663.5560  & $F=2-1$		& 4 & -2$\to$2 & 2.04\\
H$^{13}$CO$^+$(1--0)	& 86754.2884 & N/A		& 5\tablenotemark{e} &
							    -2$\to$2 &  2.02 \\
HCO$^+$(1--0)		& 89188.5260  & N/A		& 6 & -5$\to$5 & 3.23 \\
SiO(2--1)		& 86846.9600  & N/A		& 7 & -1.5$\to$1.5 & 1.76 \\
NH$_2$D(1,1)		& 85926.2630  & N/A		& 8 & -6.5$\to$6.5 & 3.66 \\
\enddata
\tablenotetext{a}{In the case of a species with hyperfine splitting,
the transition used for centering.}
\tablenotetext{b}{Reference for the adopted frequency:  1.~\citet{Pagani09}; 
	2.~\citet{Pearson76}, as recommended by JPL 
	({\tt http://spec.jpl.nasa.gov}); 
	3.~\citet{Ahrens02}, as recommended by CDMS \citep{Muller05};
	4.~\citet{Bechtel06};
	5.~\citet{Schmid-Burgk04}, as recommended by CDMS;
	6.~\citet{Ulich76}, as recommended by NIST \citep{Lovas09}
	7.~\citet{Manson77}, as recommended by CDMS;
	8.~Main frequency listed for no hyperfine splitting from 
	 \citet{Bester83}, as recommended by NIST (our sensitivity
	is too low to pick up the satellite components).}
\tablenotetext{c}{The range in velocities, relative to the line centroid,
	used when calculating the integrated intensity.}
\tablenotetext{d}{The median formal 1-sigma error on the integrated intensity
	in brightness temperature units,
	i.e., the typical spectral rms multipled by the square root of the
	number of spectral channels integrated over.}
\tablenotetext{e}{Note that hyperfine splitting was detected by 
	{Schmid-Burgk04}, but with the components separated by only 
	0.133~km~s$^{-1}$, this would not  be visible in our Mopra data.}
\end{deluxetable}

\begin{deluxetable}{cccccc}
\tabletypesize{\normalsize}
\tablecolumns{6}
\tablecaption{Hill5 Model Parameters Fit to HNC \label{tab_HNC_Hill5} }
\tablehead{
\colhead{Fit\tablenotemark{a}}&
\colhead{$V_{infall}$\tablenotemark{b}} &
\colhead{$\delta V$\tablenotemark{b}} &
\colhead{$V_{lsr}$\tablenotemark{b}} &
\colhead{$\tau$\tablenotemark{b}} &
\colhead{$T_{peak}$\tablenotemark{b}} \\
\colhead{ } &
\colhead{(km~s~$^{-1}$)} &
\colhead{(km~s~$^{-1}$)} &
\colhead{(km~s~$^{-1}$)} &
\colhead{ } &
\colhead{(K)} 
}
\startdata
Fit-1& 0.25$\pm$0.02&0.45 $\pm$0.01 &0.05$\pm$0.01 &2.7$\pm$0.1 &6.33$\pm$0.06\\
Fit-2& 0.54$\pm$0.01 &0.36 $\pm$0.01 &0.06$\pm$0.01 &1.6$\pm$0.1 &6.6$\pm$0.1\\
\enddata
\tablenotetext{a}{In our 1000 randomized fits, two clusters of fit parameters
	were identified, as listed here.  The original HNC fitted Hill5 
	parameters are all identical to the mean values listed for Fit-1.}
\tablenotetext{b}{Mean and standard deviations of the infall velocity,
	velocity dispersion (Gaussian sigma units), centroid velocity,
	optical depth, and peak excitation temperature, for the subset of the
	1000 randomized trials.}
\end{deluxetable}

\end{document}